\documentclass[onecollarge,natbib]{svjour2}
\bibpunct{[}{]}{;}{n}{}{,} % to get "[numbered]" references from natbib
\smartqed  % flush right qed marks, e.g. at end of proof
\usepackage{graphicx}
\usepackage{amsmath}
\usepackage{bm}
\usepackage{notoccite}
\usepackage{amsmath}
\usepackage{multirow}
\journalname{Few-Body Systems}
\begin{document}

\title{Light-front holography in $B$ physics %\thanks{Grants or other notes
}


\author{M. Ahmady    \and S. Lord \and
R. Sandapen
}


\institute{R. Sandapen \at
              Universit\'e de Moncton $\&$ Mount Allison University\\
              \email{ruben.sandapen@umoncton.ca}           %  \\
             \and
            M. Ahmady \at
            Mount Allison University\\
           \email{mahmady@mta.ca}
            \and
            S. Lord \at
            Universit\'e de Moncton\\
           \email{esl8420@umoncton.ca}
}

\date{Received: date / Accepted: date}

\maketitle

\begin{abstract}
We use AdS/QCD holographic Distributions Amplitudes for the vector mesons $\rho$ and $K^*$ in order to predict the radiative decays $B \to \rho \gamma$ and $B \to K^* \gamma$  as well as the semileptonic decays $B \to \rho l \nu$ and $B \to K^* \mu^+ \mu^-$. For the radiative decays, we find that the AdS/QCD DAs offer the advantage of avoiding end-point singularities when computing power suppressed contributions.  For the semileptonic decays $B \to \rho l \nu$ and $B \to K^* \mu^+ \mu^-$, we compare our predictions to the latest BaBar and LHCb data respectively.  
\end{abstract}

\section{Introduction}
\label{intro}
The study of exclusive $B$ decays to vector mesons is useful for precision tests of the Standard Model (SM) as well as for probing New Physics (NP) beyond the SM. For instance, the semileptonic $B \to \rho l \nu$ allows the determination of the CKM matrix element $V_{ub}$. The radiative decays $B \to (\rho,K^*) \gamma$ and the dileptonic decay $B \to K^* \mu^+ \mu^-$ are rare flavor-changing neutral current which are suppressed in the SM and thus sensitive to NP contributions. The computation of observables for these decays depend explicitly or implicitly (via form factors) on the Distribution Amplitudes of the vector mesons. Traditionally, these DAs are expressed as truncated Gegenbauer polynomials whose moments are fixed using QCD Sum Rules \cite{Ball:1998ff}. We use here an alternative method whereby the DAs are given in terms of the light-front wavefunction of the vector meson with the wavefunction itself being obtained using light-front holography \cite{deTeramond:2008ht}. For a recent review of light-front holography, we refer to \cite{Brodsky:2014yha}.

In light-front QCD, with massless quarks, the meson wavefunction can be written in the following factorized form\cite{deTeramond:2008ht}: 
 \begin{equation}
 \phi(z,\zeta, \varphi)=\frac{\Phi(\zeta)}{\sqrt{2\pi \zeta}} f(z) \mathrm{e}^{i L \varphi}  
 \label{factorized-lc}
 \end{equation}
with $\Phi(\zeta)$ satisfying the so-called holographic light-front Schroedinger equation 
\begin{equation}
\left(-\frac{d^{2}}{d\zeta^2}-\frac{1-4L^{2}}{4\zeta^{2}}+U(\zeta)\right)\Phi(\zeta)=M^{2}\Phi(\zeta)
\label{hLFSE} 
\end{equation}
and where $L$ is the orbital angular momentum quantum number, $z$ is the fraction of the meson light-front momentum carried by the quark and the variable $\zeta=\sqrt{z(1-z)}r$ where $r$ is the transverse distance between the quark and the antiquark. Remarkably, if $\zeta$ is identified with the fifth dimension of AdS space, then equation \eqref{hLFSE} maps onto the wave equation for spin-$J$ string modes in a modified anti-de Sitter space. Modifying the geometry of AdS with a quadratic dilaton field then yields the confining potential in physical spacetime as 
\begin{equation}
 U(\zeta)=\kappa^4 \zeta^2 + 2\kappa^2(J-1) 
\label{quadratic-dilaton}
\end{equation}
The holographic light-front wavefunction for a vector meson $(L=0,S=1)$ then becomes

\begin{equation}
\phi_{\lambda} (z,\zeta) \propto \sqrt{z(1-z)} \exp
\left(-\frac{\kappa^2 \zeta^2}{2}\right)
\exp\left \{-\left[\frac{m_q^2-z(m_q^2-m^2_{\bar{q}})}{2\kappa^2 z (1-z)} \right] 
\right \} \label{AdS-QCD-wfn}
\end{equation}
with $\kappa=M_{V}/\sqrt{2}$ and where we have introduced the dependence on quark masses following a prescription by Brodsky and de T\'eramond \cite{deTeramond:2012rt}. This holographic wavefunction has been used to successfully  predict diffractive $\rho$-meson electroproduction\cite{Forshaw:2012im}. We allow the normalization of the wavefunction to depend on the polarization $\lambda$ of the vector meson; hence the subscript $\lambda$ on $\phi$.

\section{Distribution Amplitudes and radiative $B$ decays}
\label{sec:1}
Up to twist-$3$ accuracy, there are four DAs for a vector meson defined via the following equations \cite{Ball:2007zt}
\begin{eqnarray}
\langle 0|\bar q(0)  \gamma^\mu q(x^-)| V
(P,\lambda)\rangle
&=& f_{V} M_{V}
\frac{e_{\lambda} \cdot x}{P^+x^-}\, P^\mu \int_0^1 \mathrm{d} u \; e^{-iu P^+x^-}
\phi_{V}^\parallel(u,\mu)
\nonumber \\
&\hspace{-1.0cm}+&\hspace{-0.5cm} f_{V} M_{V}
\left(e_{\lambda}^\mu-P^\mu\frac{e_{\lambda} \cdot
x}{P^+x^-}\right)
\int_0^1 \mathrm{d} u \; e^{-iu P^+x^-} g^{\perp (v)}_{V}(u,\mu)  \;,
\label{DA:phiparallel-gvperp}
\end{eqnarray}
\begin{equation}
\langle 0|\bar q(0) [\gamma^\mu,\gamma^\nu] q (x^-)|V
(P,\lambda)\rangle =2 f_{V}^{\perp} (e^{\mu}_{\lambda} P^{\nu} -
e^{\nu}_{\lambda} P^{\mu}) \int_0^1 \mathrm{d} u \; e^{-iuP^+ x^-} \phi_{V}^{\perp}
(u, \mu) \label{DA:phiperp}
\end{equation}
and
\begin{equation}
\langle 0|\bar q(0) \gamma^\mu \gamma^5 s(x^-)|V (P,\lambda)\rangle
=-\frac{1}{4} \epsilon^{\mu}_{\nu\rho\sigma} e_{\lambda}^{\nu}
P^{\rho} x^{\sigma}  \tilde{f}_{V} M_{V} \int_0^1 \mathrm{d} u \; e^{-iuP^+ x^-}
g_{V}^{\perp (a)}(u, \mu) \label{DA:gaperp}
\end{equation}
where 
\begin{equation}
\tilde{f}_{V} = f_{V}-f_{V}^{\perp} \left(\frac{m_q + m_{\bar{q}}}{M_{K^*}} \right) \;.
\end{equation}
We have previously shown that \cite{Ahmady:2013cva}
\begin{equation}
\phi_{V}^\parallel(z,\mu) =\frac{N_c}{\pi f_{V} M_{V}} \int \mathrm{d}
r \mu
J_1(\mu r) [M_{V}^2 z(1-z) + m_{\bar{q}} m_{q} -\nabla_r^2] \frac{\phi_{L}(r,z)}{z(1-z)} \;,
\label{phiparallel-phiL}
\end{equation}
\begin{equation}
\phi_{V}^\perp(z,\mu) =\frac{N_c }{\pi f_{V}^{\perp}} \int \mathrm{d}
r \mu
J_1(\mu r) [m_q - z(m_q-m_{\bar{q}})] \frac{\phi_{T}(r,z)}{z(1-z)} \;,
\label{phiperp-phiT}
\end{equation}

\begin{equation}
g_{V}^{\perp(v)}(z,\mu)=\frac{N_c}{2 \pi f_{V} M_{V}} \int \mathrm{d} r \mu
J_1(\mu r)
\left[ (m_q - z(m_q-m_{\bar{q}}))^2 - (z^2+(1-z)^2) \nabla_r^2 \right] \frac{\phi_{T}(r,z)}{z^2 (1-z)^2
}
\label{gvperp-phiT}
\end{equation}
and
\begin{equation}
\frac{\mathrm{d} g_{V}^{\perp(a)}}{\mathrm{d} z}(z,\mu)=\frac{\sqrt{2} N_c}{\pi \tilde{f}_{V} M_{V}} \int \mathrm{d} r \mu
J_1(\mu r)
[(1-2z)(m_q^2- \nabla_r^2) + z^2(m_q+m_{\bar{q}})(m_q-m_{\bar{q}})]\frac{\phi_{T}(r,z)}{z^2 (1-z)^2} 
\label{gaperp-phiT}
\end{equation}
where $\phi_{\lambda}(r,z)$ is the holographic AdS/QCD wavefunction given by equation \eqref{AdS-QCD-wfn}. 

In Ref. \cite{Ahmady:2012dy}, we have used the AdS/QCD DAs for the $\rho$ meson to compute, beyond leading power accuracy in the heavy quark limit, the branching ratio for the radiative decay $B \to \rho \gamma$ as well as the branching ratio for the very rare power-suppressed decay $B_s \to \rho \gamma$. Furthermore, in Ref. \cite{Ahmady:2013cva}, we have used the AdS/QCD DAs for the $K^*$ in order to compute the isospin asymmetry in the decay $B \to K^* \gamma$. 

We found that an advantage of using the AdS/QCD DAs is that end-point singularities are avoided when computing some power-suppressed contributions \cite{Ahmady:2012dy,Ahmady:2013cva}. For example, the integral 
\begin{equation}
X_{\perp}( \mu_h) = \int_0^1 \mathrm{d} z \; \phi_{K^*}^{\perp}(z, \mu_h) \left( \frac{1 + \bar{z}}{3 \bar{z}^2} \right)
\label{Xperp}
\end{equation}
which appears in the computation of the isospin asymmetry in $B \to K^* \gamma$ does not diverge with the holographic twist-$2$ DA given by equation \eqref{phiperp-phiT} while it suffers from an end-point singularity if the DA is assumed to be a truncated Gegenbauer polynomial.

\section{$B \to V$ transition form factors}
\label{sec:2}
In order to compute the branching ratio for the semileptonic decays $B \to \rho l \nu$ and $B \to K^* \mu^+ \mu^-$, we need to compute non-perturbative $B \to V$ transistion form factors 
as a function of the momentum transfer $q$ to the lepton pair. We use here the light-cone Sum Rules (LCSR) method \cite{Ball:1997rj,Aliev:1996hb} which requires as non-perturbative inputs the DAs of the vector mesons. We work to twist-$2$ accuracy so that only the twist-$2$ DAs given by equations \eqref{phiparallel-phiL} and \eqref{phiperp-phiT} are required. Our results for the three relevant form factors $A_1$, $A_2$ and $V$ at $q^2=0$ are collected in table \ref{tab:semiFF} and compared with existing predictions in the literature. 

\begin{table}[t]
\caption{Our predictions, corresponding to $m_f=0.05,0.14,0.35$ GeV, for the semileptonic form factors compared to the sum rules predictions of Ref. \cite{Ball:1997rj} and the quark model predictions of Ref.  \cite{Faustov:1995bf,Wirbel:1985ji,Jaus:1989au,Melikhov:1996pr}.}
\centering
\label{tab:semiFF}       % Give a unique label
\begin{tabular}{lllllll}
\hline\noalign{\smallskip}
{Form factor}  & {AdS/QCD} & {BB} & {FGM} & {WSB} & {Jaus}  & {Melikhov}  \\[3pt]
\tableheadseprule\noalign{\smallskip}
$A_1(0)$ & $0.17$, $0.25$, $0.25$ & $0.27 \pm 0.05$ & $0.26 \pm 0.03$ & $0.28$ & $0.26$ & $0.17$ - $0.18$ \\
$A_2(0)$ & $0.15$, $0.26$, $0.27$ & $0.28 \pm 0.05$ & $0.31 \pm 0.03$ & $0.28$ & $0.24$ & $0.155$ \\
$V(0)$ & $0.23$, $0.33$, $0.32$ & $0.35 \pm 0.07$ & $0.29 \pm 0.03$ & $0.33$ & $0.35$ & $0.215$ \\
\noalign{\smallskip}\hline
\end{tabular}
\end{table}

Having computed the form factors, we are able to compute $V_{ub}$-independent observables like the ratio of partial decay widths in various $q^2$ bins. The BaBar collaboration has measured partial decay widths in three different $q^2$ bins:\cite{delAmoSanchez:2010af}
\begin{equation}
\Gamma_{\mbox{\tiny{low}}}= \int_0^8 \frac{\mathrm{d} \Gamma}{\mathrm{d} q^2} \mathrm{d} q^2 = (0.564 \pm 0.166) \times 10^{-4} 
\end{equation}
for the low $q^2$ bin,
\begin{equation}
\Gamma_{\mbox{\tiny{mid}}}= \int_8^{16} \frac{\mathrm{d} \Gamma}{\mathrm{d} q^2} \mathrm{d} q^2 = (0.912 \pm 0.147) \times 10^{-4} 
\end{equation}
for the intermediate $q^2$ bin and
\begin{equation}
\Gamma_{\mbox{\tiny{high}}}= \int_{16}^{20.3} \frac{\mathrm{d} \Gamma}{\mathrm{d} q^2} \mathrm{d} q^2 = (0.268 \pm 0.062) \times 10^{-4} 
\end{equation}
for the high $q^2$ bin. 
From these measurements, we can thus deduce the $|V_{ub}|$-independent ratios of partial decay widths
\begin{equation}
R_{\mbox{\tiny{low}}}=\frac{\Gamma_{\mbox{\tiny{low}}}}{\Gamma_{\mbox{\tiny{mid}}}}=0.618 \pm 0.207
\end{equation}
and 
\begin{equation}
R_{\mbox{\tiny{high}}}=\frac{\Gamma_{\mbox{\tiny{high}}}}{\Gamma_{\mbox{\tiny{mid}}}}=0.294 \pm 0.083
\end{equation}
which we compare to our predictions: $R_{\mbox{\tiny{low}}}=0.580, 0.424$, $R_{\mbox{\tiny{high}}}=0.427,0.503$ for $m_q=0.14, 0.35$ GeV respectively.  Our predictions for $R_{\mbox{\tiny{low}}}$ are therefore in agreement with the BaBar measurement. This is not the case for $R_{\mbox{\tiny{high}}}$ where our predictions are above the BaBar measurement. This is perhaps not unexpected given that the LCSR predictions are less reliable in the high $q^2$ bin. 
\begin{figure}
\centering
  \includegraphics[width=0.50\textwidth]{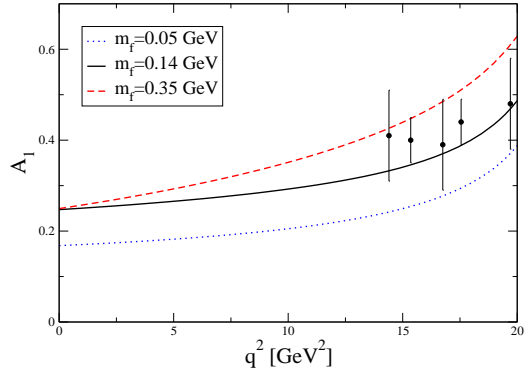}
\caption{The semileptonic $B \to \rho$ form factor $A_1$ of a function of the momentum transfer squared $q^2$.We extrapolate our predictions to high $q^2$ in order to compare to the lattice data from the UKQCD collaboration\cite{Flynn:1996rc,DelDebbio:1997kr,Burford:1995fc}}
\label{fig:A1}       % Give a unique label
\end{figure}

\begin{figure*}
\centering
  \includegraphics[width=0.50\textwidth]{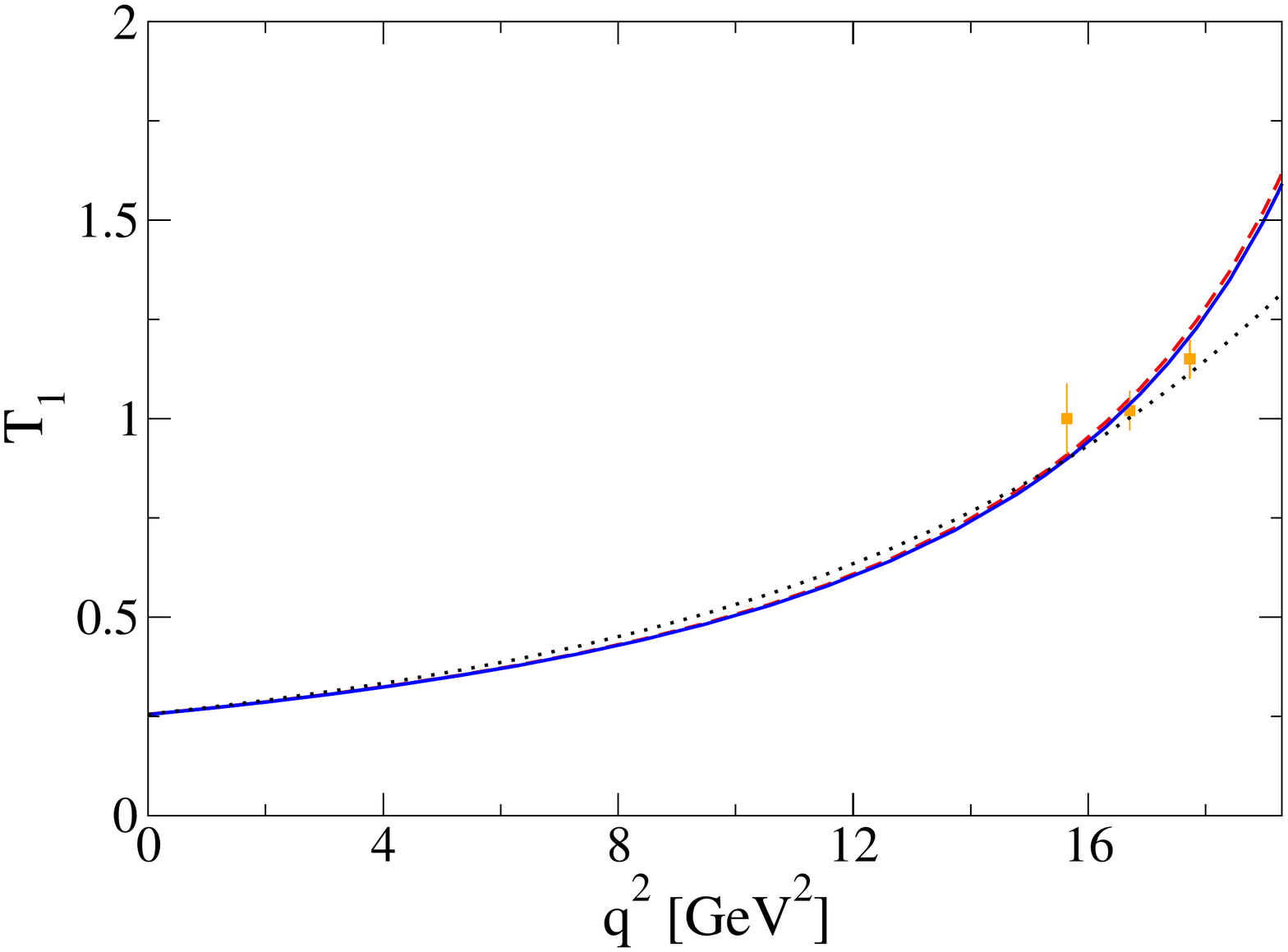}\includegraphics[width=0.50\textwidth]{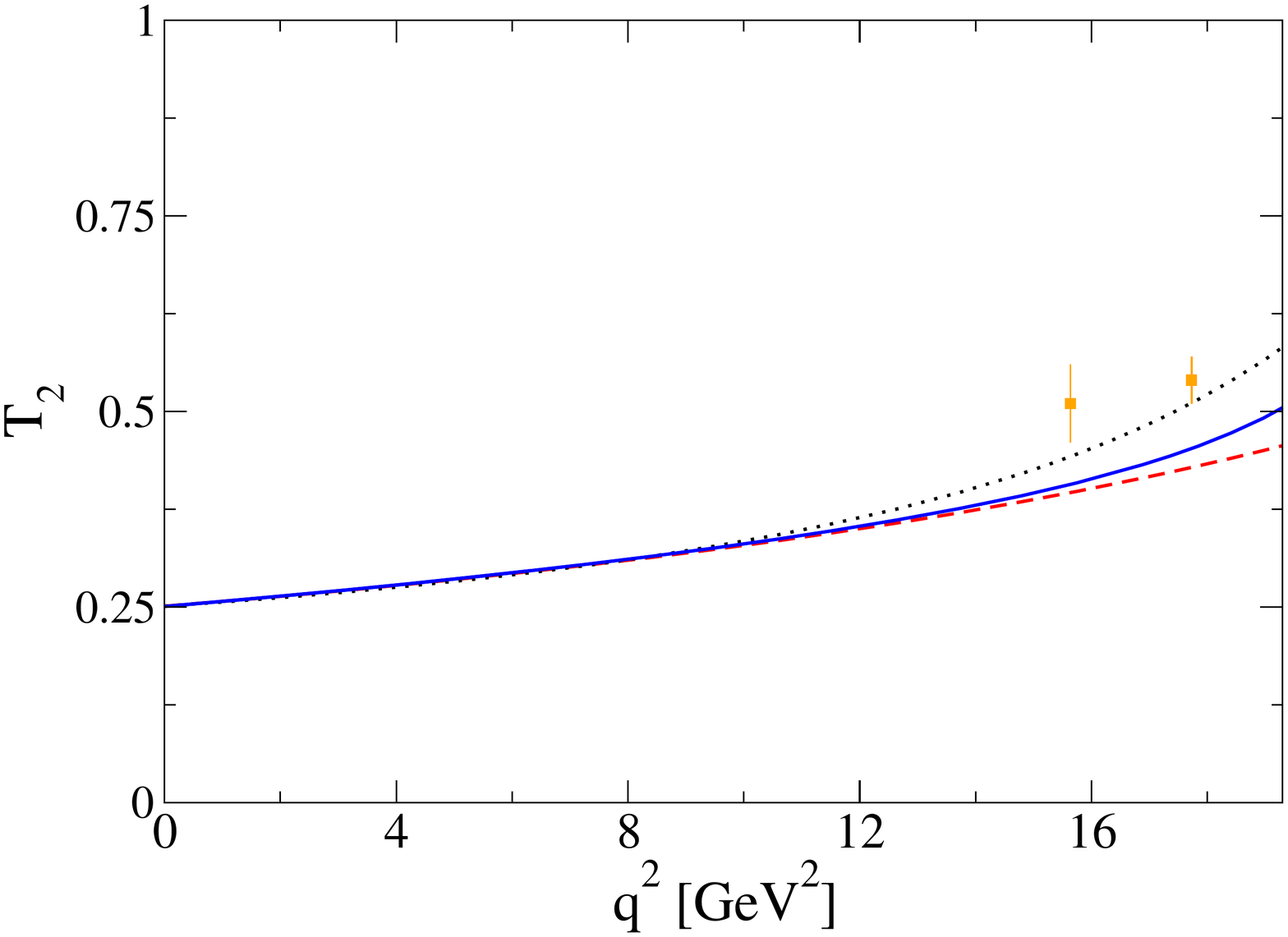}
\caption{Two of the seven $B\to K^*$ transition form factors. The lattice data is from Ref. \cite{Horgan:2013hoa}. AdS/QCD: solid blue, AdS/QCD $+$ lattice : dotted black. AdS/QCD fit: dashed red.}
\label{fig:FFKstar}       % Give a unique label
\end{figure*}

For the $B \to K^*$ transition, there are seven relevant form factors. Again we compute them using LCSR with the $K^*$ AdS/QCD holographic DAs as input. Our results for two of the three tensor form factors $T_1$ and $T_2$ are shown in Figure \ref{fig:FFKstar}. The results for the full set of form factors can be found in Ref. \cite{Ahmady:2014sva}. The solid blue curves are the AdS/QCD-LCSR predictions extrapolated to high $q^2$. We also do  fits using the form
\begin{equation}
F(q^2)=\frac{F(0)}{1- a (q^2/m_B^2) + b (q^4/m_B^4)}
\label{FitFF}
\end{equation}
to the AdS/QCD predictions in the region of reliability of the LCSR and extrapolate to high $q^2$. These are the red dashed curves. Finally, we include in our fits the lattice data \cite{Horgan:2013hoa} available at large $q^2$ to generate the dashed black curves.  
% Confirm this citation! - SL
% I'm fairly certain it is the right one (it is the one we citted for Lattice data in our first B->K paper. - SL

\begin{table}[t]
\caption{The values of the form factors at $q^2=0$ together with the fitted parameters $a$ and $b$. The values of $a$ and $b$ are obtained by fitting Eq. \eqref{FitFF} to either the AdS/QCD predictions for low-to-intermediate $q^2$ or both the AdS/QCD predictions for low-to-intermediate $q^2$ and the lattice data at high $q^2$.}
\centering
\label{tab:abAdS}       % Give a unique label
\begin{tabular}{llllll}
\hline\noalign{\smallskip}
 & & \multicolumn{2}{l}{AdS/QCD} & \multicolumn{2}{l}{AdS/QCD + Lat.} \\ \cline{3-4} \cline{5-6}
F & $F(0)$ & $a$ & $b$ & $a$ & $b$ \\[3pt]
\tableheadseprule\noalign{\smallskip}
$A_0$ & $0.285$ & $1.158$ & $0.096$ & $1.314$ & $0.160$ \\
$A_1$ & $0.249$ & $0.625$ & $-0.119$ & $0.537$ & $-0.403$ \\
$A_2$ & $0.235$ & $1.438$ & $0.554$ & $1.895$ & $1.453$ \\
$V$ & $0.277$ & $1.642$ & $0.600$ & $1.783$ & $0.840$ \\
$T_1$ & $0.255$ & $1.557$ & $0.499$ & $1.750$ & $0.842$ \\
$T_2$ & $0.251$ & $0.665$ & $-0.028$ & $0.555$ & $-0.379$  \\
$T_3$ & $0.155$ & $1.503$ & $0.695$ & $1.208$ & $-0.030$ \\
\noalign{\smallskip}\hline
\end{tabular}
\end{table}

\begin{figure*}
\centering
  \includegraphics[width=0.75\textwidth]{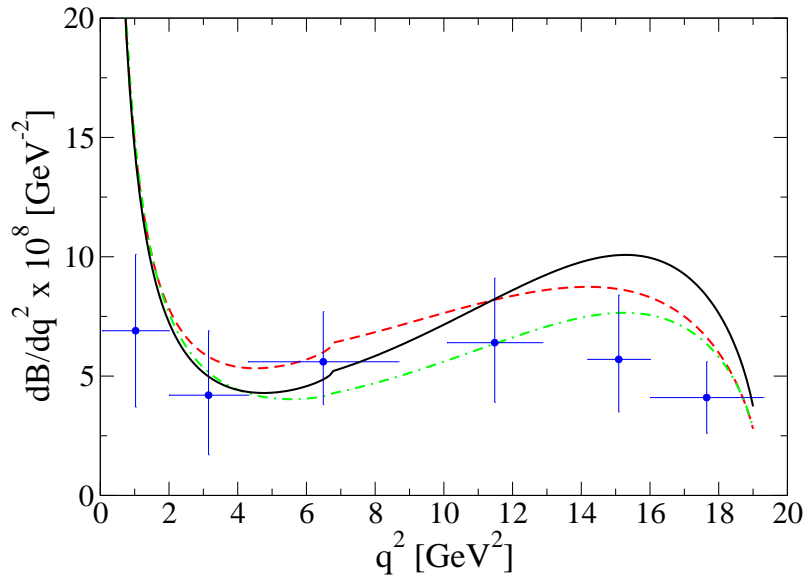}
\caption{The differential branching ratio of the $B \to K^* \mu^+ \mu^-$ process as a function of $q^2$. The dashed red line is the AdS/QCD result, solid black is the AdS/QCD+lattice fit and the dot-dash green curve is the AdS/QCD+lattice+NP result. Note that we average the LHCb experimental result data from both the $B^{\circ} \to K^{*\circ} \mu^+ \mu^-$ \cite{LHCb-CONF-2012-008} and $B^{+} \to K^{*+} \mu^+ \mu^-$ \cite{Aaij:2012cq} data.}
\label{fig:BR}       % Give a unique label
\end{figure*}

Finally, we use the $B \to K^*$ form factors to compute the differential branching ratio for $B \to K^* \mu^+ \mu^-$. Our results are shown in figure \ref{fig:BR}. As can be seen, our AdS/QCD prediction (dashed red curve) tend to overshoot the data at high $q^2$. Using the form factors fitted to the lattice data does not remedy the situation: see the black solid curve. On the other hand, we are able to achieve agreement at high $q^2$ by adding a new physics contribution to the Wilson coefficient $C_9$ \cite{Descotes-Genon:2013vna}.   

%

%\begin{table}[t]
%\centering
%\begin{tabular}{lll}
%\hline\noalign{\smallskip}
%first & second & third  \\[3pt]
%\tableheadseprule\noalign{\smallskip}
%\end{tabular}
%\end{table}
\section{Conclusion}
Light-front holography provides a new alternative method for computing the Distribution Amplitudes of light mesons. We have investigated the use of holographic Distribution Amplitudes in computing observables for radiative and semileptonic B decays to light vector mesons. It would be interesting to extend our study to B decays to pseudoscalar mesons. 

\begin{acknowledgements}
This research is supported by the Natural Sciences and Engineering Research Council of Canada (NSERC). R.S. thanks the organizers of light-cone 2014 for their invitation and a very enjoyable workshop at NCSU.
\end{acknowledgements}

% BibTeX users please use
\bibliographystyle{spbasic.bst}
\bibliography{SandapenLC2014}   % name your BibTeX data base

%\begin{thebibliography}{3}
%\bibitem[Author I(1999)]{Ref1}
%Author I (year) Article title. Journal Title-Abbreviated Vol: pp--pp
%\bibitem[Author and Smith(2001)]{Ref2}
%Author I, Smith J (year) Book title. Publisher, Place, pp numbers
% Format for proceedings
%\bibitem[Author and Smith(2003)]{Ref3}
%Author I, Smith J (year) Paper title. In: Editor, A. (ed.) Proceedings
%Title, Location, Date, pages. Publisher, Place
% etc
%\end{thebibliography}

\end{document}